\documentclass[12pt]{JHEP3}
\usepackage{epsfig}
\preprint{ {\tt hep-th/0208011} }
\newcommand{\be}[1]{ \begin{equation}\label{#1} }
\newcommand{\ee}{\end{equation}}
\newcommand{\bea}[1]{\begin{eqnarray}\label{#1} }
\newcommand{\eea}{\end{eqnarray}}
\newcommand{\eq}[1]{(\ref{#1})}

\title{Unstable magnetic fluxes in heterotic string theory}
\author{Justin R. David \\ Department of Physics, University of
California,\\Santa Barbara, CA 93106, USA.\\
\email{justin@vulcan.physics.ucsb.edu} }
\abstract{ We study 
exact string backgrounds representing
a constant magnetic field background in heterotic string theory.
These backgrounds are obtained by Kaluza-Klein reduction of a
special class of plane wave solutions.
For small values of the magnetic field 
they possess localized closed string
tachyons analogous to the Nielsen-Olesen instability of a constant
magnetic field in $SO(3)$ Yang-Mills theory. 
When the magnetic field is embedded in the $SO(32)$ gauge group of the
heterotic string it is possible to
study the lowest level tachyon in supergravity.
We identify the closed
string tachyons as  fluctuations of the supergravity fields about the
background. We argue that the tachyon signals the decay of the
background to flat space. Our evidence rests on the study of the
closed string tachyon potential, world sheet renormalization group
equations in the supergravity approximation
and the S-dual of this system in type I theory.
In the S-dual
description, the closed string tachyons in heterotic
string theory correspond to open string tachyons in type I theory.
Finally we analyze the non-perturbative stability of 
these models representing constant magnetic field backgrounds 
and  show that it is stable to pair production of particles.}
\keywords{Tachyon condensation, Heterotic string theory}
\begin{document}
\baselineskip 3.5ex
\section{Introduction}

Open string tachyon condensation has been studied in a great detail
following the work of Sen \cite{Sen:1998rg}. Open string tachyons
occur in various non-supersymmetric configurations of D-branes. The
tachyons are relevant operators on the boundary of the open string
world sheet which  modify the boundary conditions of the open string. 
The end point of
tachyon condensation for the various systems studied is flat space or
a supersymmetric brane configuration. For example the end point of 
condensation of the tachyonic mode when a D-brane and anti D-brane are
brought on top of each other is flat space. In fact the minimum of the
tachyon potential equals the tension of the D-branes. 

On the other hand tachyons in closed string theory are relevant
operators of the bulk  world sheet theory. Condensation of these modes
can potentially change the space time in 
which the string propagates.  
For example, it is well known that 
relevant operators in the bulk conformal field
theory change its central charge \cite{Zamolodchikov:1986gt}. 
At present we lack tractable methods to study bulk tachyons. 
There has been progress in understanding condensation of localized
closed string tachyons 
\cite{Adams:2001sv,Dabholkar:2001gz,Harvey:2001wm,
Vafa:2001ra,Dabholkar:2001wn,Russo:2001tf,David:2001vm}.
These tachyons occur in various non-supersymmetric orbifolds 
in the twisted sector. The
flow induced by these operators has been studied by various methods.
It has been seen that condensation of  
these tachyons resolves the orbifold and the system flows to flat
space.

Localized closed string tachyons are similar to open string tachyons,
they are localized on defects in the space time, just as open string
tachyons are localized on D-branes. It would be interesting to make
this connection precise, by finding
an example of a system in which the tachyonic
mode can have both an open and closed string description. 
The existence of an open string dual also suggests that the closed
string tachyon potential can be evaluated.  
In this paper we study such a system. 
These systems were first discovered by
\cite{Horowitz:1995rf,Kiritsis:1995iu} 
and studied in detail by \cite{Russo:1995cv,Russo:1995aj}. 
They are backgrounds representing uniform magnetic fields in string
theory.

To motivate these
backgrounds we first discuss a well known instability in Yang-Mills
theory.
It is known that a constant magnetic field in $SO(3)$ Yang-Mills theory
is unstable \cite{Nielsen:1978rm}. This instability is called the
Nielsen-Olesen instability, 
there is an infinite degeneracy of
tachyons at the lowest Landau level. 
The tachyon potential can be evaluated and it can be
shown that the value of the potential at the minimum cancels the
energy of the magnetic flux. Thus these tachyons signal the instability
of the constant magnetic field to decay to the vacuum.

There are two known
ways to embed magnetic fields in string theory. 
They are the Melvin backgrounds 
\cite{Melvin:1964qx} and backgrounds obtained by Kaluza-Klein
reduction of a special class of 
plane wave solutions 
found by \cite{Horowitz:1995rf} called chiral null models. 
In this paper we will study the latter.
They are exact conformal field theories to all orders in
$\alpha'$. If the magnetic field is uniform they can be quantized in the
light cone gauge
\cite{Russo:1995cv,Russo:1995aj}
When these
models are embedded in type II string theories they are
supersymmetric. In heterotic string theory 
there are embeddings which have  tachyons whose
(mass)$^2$ 
agrees with the Nielsen-Olesen instability 
for weak magnetic fields. There are
two distinct situations which are related by a T-duality; 
(1) the magnetic field arises from a
Kaluza-Klein gauge field, (2) the magnetic field is
present in the $SO(32)$ gauge group 
\footnote{We choose to discuss the $SO(32)$
heterotic string theory for definiteness, a similar discussion can be
done in the $E_8\times E_8$ heterotic string.} 
of the heterotic string theory.
To study tachyon condensation it is convenient to examine the latter.
We identify the tachyon corresponding to the Nielsen-Olesen 
instability as a supergravity fluctuation.
We argue that the tachyon condensation in this background drives it to
flat space. We present three arguments in favour of this.

Firstly  we discuss the S-dual of this background in type I theory.
In the type I background closed string tachyons of the heterotic string
background appear in the open string sector. We derive a decoupling
limit in the type I background in which the back reaction to the
geometry can be neglected. 
In the decoupling limit the type I open string 
can be quantized and the open string tachyon 
corresponds to the Nielsen-Olesen instability.
Furthermore, in the decoupling
limit the process of tachyon condensation is identical to the 
condensation of the Nielsen-Olesen tachyon in field theory.
The tachyon drives the system to the Yang-Mills vacuum. Thus,
turning on the coupling, by continuity we would expect that the tachyon
in the heterotic string also drives the system to the vacuum. 
Next we analyze the tachyon condensation process directly in the
heterotic string. We construct an energy functional and derive the
closed string tachyon potential. 
We show that the minimum of the tachyon potential
cancels the background magnetic field and the system is driven to flat
space. 
In the various non-supersymmetric orbifolds studied so far 
there has been only an indirect evaluation of the closed string 
tachyon potential in \cite{Dabholkar:2001wn}. This model admits a
direct evaluation of the tachyon potential.
Finally we study the world sheet renormalization group flow
in supergravity. 
We show that it is possible to obtain a consistent set of flow
equations for small magnetic fields. 
It is shown that the tachyon drives the RG flow to 
flat space.

Background fields in a theory also can decay by pair production of
particles. For example constant electric fields decay by pair
production of electrons and positrons \cite{Schwinger:1951nm}, the
Melvin background contains a magnetic field which can
decay by pair production of Kaluza-Klein monopoles 
\cite{Dowker:1994up,Dowker:1995gb,
Dowker:1996sg}. It is of interest to know if the backgrounds studied in
this paper are unstable to pair production of particles.
For this it is convenient to study the case when the
magnetic field in heterotic string theory arises from a Kaluza-Klein
gauge field. 
By examining the corrections to the effective action
describing vacuum polarization we show
the amplitude for pair production of particles vanishes.

The organization of this paper is as follows. In section 2 we review
the Nielsen-Olesen instability and the evaluation of the tachyon
potential. In section 3 we introduce the plane wave solutions which
represent constant magnetic field backgrounds in string theory. 
Section 4 introduces the background in heterotic string theory in
which the magnetic field arises from the $SO(32)$ gauge group. We
identify the tachyons as supergravity fluctuations.
In section 5 we argue that  tachyon
condensation in this background drives it to flat space.
In section 6 we 
show that constant magnetic fields in heterotic string theory are
stable to decay by pair production of particles.

\section{The Nielsen-Olesen instability}

In this section we briefly review the Nielsen-Olesen instability in
non-Abelian gauge theories \cite{Nielsen:1978rm}. 
We first discuss the linearized
fluctuations around a constant magnetic field in $SO(3)$ Yang-Mills
theory in $3+1$  dimensions and show that there are unstable modes in
the lowest Landau level. We choose the gauge group to be $SO(3)$ as it
can be embedded trivially in the $SO(32)$ gauge group of the
heterotic string. 
Then we evaluate the tachyon potential and
show that the instability corresponds to the decay of the magnetic
field to the Yang-Mills vacuum.

\subsection{Fluctuations around a constant magnetic field}

The action is given by of $SO(3)$ Yang-Mills theory in $3+1$
dimensions is given by
\be{YMaction}
S_{\rm{YM}}= -\frac{1}{4 g_{\rm{YM}}^2 }  
\int d^4 x \rm{Tr} (F_{\mu\nu}
F^{\mu\nu}).
\ee
Here the $SO(3)$ gauge group generators are given by following
$3\times 3$ anti-symmetric matrices
\be{generators}
L_1 =
\left(
\begin{array}{ccc}
0 & 0& 0 \\
0 & 0& -i \\
0 & i &0 
\end{array}
\right), \quad
L_2 =
\left(
\begin{array}{ccc}
0 & 0& i \\
0 & 0& 0 \\
-i & 0 & 0
\end{array}
\right), \quad
L_3= 
\left(
\begin{array}{ccc}
0 & -i & 0 \\
i & 0 &0  \\
0 & 0& 0 
\end{array}
\right). \quad
\ee
They obey the usual $SO(3)$ commutation relations $[L_i, L_j] =
i\epsilon_{ijk}L_k$. 
We choose the constant magnetic field to be along the $3$ direction  
and in the $3$ direction of the gauge group. 
The field configuration in the Landau gauge is given by
\be{background}
A_0 = 0, \quad A_1 = Qf x^2  
L_3, \quad A_2 = 0,
 \quad A_4 = 0.
\ee
Without loss of generality we can take $Qf>0$ \footnote{For gauge
fields in $SO(32)$ heterotic string theory $Q=\sqrt{2}$.}.

We now now analyze the
fluctuations around this background and show that there are unstable
modes in the lowest Landau level. 
A general fluctuation can be
written as
\be{fluctuation}
\delta A_\mu = W_\mu^+ L_+ + W_\mu^- L_- + \phi_\mu  L_3,
\ee
where $L_+ = L_1 +iL_2$ and $L_- = L_1 -iL_2$.
The linearized equation of motion obeyed by the fluctuation is given by
\be{fluct-eom}
D^\mu D_\mu \delta A_\nu 
-2i[F^\mu\,_\nu, \delta A_\mu] = 0,
\ee
where $D_\mu$ refers to the covariant derivative given by
$D_\mu = \partial_\mu + i[A_\mu, \;  ]$. $A_\mu$ and $F_{\mu\nu}$
are the background gauge fields. 
In the above equation we have imposed the background gauge condition
\be{back-gauge}
D^\mu \delta A_\mu =0.
\ee
Consider the off diagonal fluctuation $W^+_\mu$ with 
$W^+_0 = W^+_3 =0$. The linearized equation of motion for this
fluctuation reduces to 
\be{red-eom}
\left( 
\partial^0\partial_0 + \partial^3\partial_3 + (\partial_1 - i fQx^2)^2
+ \partial_2^2
\right)W^+_{a}  - 2i \epsilon_{ab}fQ W^+_b =0,
\ee
here $a, b= 1,2$ and $\epsilon_{12} = -\epsilon_{21} =1$.
It is now clear that the eigenvalue problem reduces to the familiar
Landau level problem in quantum mechanics.
From \eq{red-eom} it can be seen that we can label 
the eigen functions by energy, $E$ and the momentum along $1$ and $3$
directions which we call $k_1$ and $k_3$ respectively. Using these
eigen values we obtain the following eigen value equation for the
energy
\be{red-eom1}
\left(
E^2 -k_3^2 - (k_1 - fQx^2)^2  + \partial_2^2 \right) 
W^+_a - 2i\epsilon_{ab} fQW^+_b =0.
\ee
The energy eigen values are given by
\be{energy}
E^2 = k_3^2 + 2fQ(l + 1/2) \mp 2fQ.
\ee
The last term arises from the fact that $i\epsilon_{ab}$ acts as a Pauli
matrix $\sigma^2$ 
with eigen values $\pm 1$, $l$ labels the Landau level. 
It is now easy to see that for $k_3
=0, l =0$ the energy is complex which signals a tachyon,  
the wave function of this tachyon is given by.
\be{wavefn}
W_a^+ (k_1) = \left(\frac{fQ}{4\pi l_1^2}\right)^{1/4} 
e^{-ik_1x^1} e^{-\frac{Qf}{2}(
-\frac{k_1}{Qf} + x^2)^2 }
\left(
\begin{array}{c}
1 \\ i
\end{array}
\right),
\ee
$E^2 = -fQ$ for this wave
function and it satisfies the background gauge condition. 
The wave function is a Gaussian and therefore these tachyons
are localized. 
We have normalized this wave function by requiring
\be{norm}
\int dx^1 dx^2 (W^+)^\dagger (k_1) W^+(k_1') = \frac{2\pi}{l_1}
\delta(k_1-k_1').
\ee
To regulate the system we confine it to a 
large box of size $l_1$ and $l_2$ in the $x$ and $y$ directions.
Note that this energy is not a function of $k_1$ and therefore 
infinitely degenerate, the degeneracy is given by 
$ fQl_1l_2/(2\pi)$. 
For completeness we discuss the tachyons from $W_a^-$,
which obey the equation
\be{red-eom2}
\left( E^2 -k_3^2 - (k_1 + fQx^2)^2  + \partial_2^2 \right) 
W^-_a + 2i\epsilon_{ab} fQW^-_b =0.
\ee
The energy eigen-values are given by
$E^2 = k_3^2 + 2fQ(l+1/2) \pm 2fQ$.
Thus again there is a tachyon in the lowest Landau level, 
the normalized tachyon wave function 
which satisfies the background gauge condition 
is given by
\be{wavefn1}
W_a^- (k_1) = \left(\frac{fQ}{4\pi l_1^2}\right)^{1/4} 
e^{-ik_1x^1} e^{-\frac{fQ}{2}(
\frac{k_1}{fQ} + x^2)^2 }
\left(
\begin{array}{c}
1 \\ -i
\end{array}
\right).
\ee
Note that the tachyon wave functions in \eq{wavefn} and \eq{wavefn1}
are related by $(W^+(k_1))^* = W^-(-k_1)$, this is true because the
gauge field is Hermitian. Higher Landau levels are obtained by
multiplying the wave functions in \eq{wavefn} and \eq{wavefn1} by the
appropriate Hermite polynomial.

Finally we analyze the diagonal fluctuation $\phi_\mu$. 
These fluctuations commute with the background given in
\eq{background} therefore they are not charged and they obey 
the Klein-Gordan equation 
\be{kgeom}
\partial^\nu\partial_\nu \phi_\mu = 0.
\ee
The wave functions satisfying the background gauge condition are given by
\be{massless}
\phi_a(k_1, k_2) = \frac{1}{\sqrt{l_1l_2}}
\frac{i\epsilon_{ab} k_b}{|k|} e^{ik_1x^1 + ik_2x^2},
\ee
with $\phi_0=\phi_3=0$ and $|k| = \sqrt{k_1^2 + k_2^2}$, 
we have set $k_3=0$. The $i$ in the above
equation is introduced so that $\phi_a^*(k) = \phi_a(-k)$.  
The normalization of these wave functions is fixed so that
\be{norm2}
\int dx^1 dx^2 \phi_a^* (k) \phi_a(k') = \frac{4\pi^2 }{l_1l_2}
\delta^2(k-k').
\ee

It is useful to state the form of the wave function 
for the off diagonal fluctuations
in the gauge for
which the background field is given by
\be{rad-back}
A_0 =0, \quad  A_i = \frac{Qf}{2} \epsilon_{ij} x^j L_3, \quad A_4=0.
\ee
The lowest Landau level wave functions is a Gaussian given by
\be{rad-wavfn}
W_a^+ = e^{-\frac{Qf}{2} \left( (x^1)^2 + (x^2)^2\right) }
\left(
\begin{array}{c}
1 \\ i
\end{array}\right), \quad W_a^- = (W_a^+)^*.
\ee
Higher Landau levels are obtained by multiplying the above wave
function with the appropriate Laguerre polynomial.

\subsection{Condensation of the Nielsen-Olesen tachyon}

We now show that the Nielsen-Olesen instability signals the decay of
the constant magnetic field in non-Abelian gauge theories to the
vacuum. We infer this using two methods.
(1) We show that for a single unit of magnetic flux
there is a tachyon potential such that the value at the minimum of the
potential cancels the energy due to the magnetic flux.
We follow the method developed in \cite{Antoniadis:1998mm} 
for evaluating the tachyon potential for uniform magnetic fields 
on a 2-torus.  We first
restrict ourselves to the tachyonic modes in the lowest Landau level
and the diagonal fluctuations. We expand the gauge field in these
modes, substitute these expansions in the action \eq{YMaction} and
eliminate the diagonal modes using the classical equations of
motion to obtain the effective potential for the tachyon. 
We then argue that the property of the tachyon potential to cancel the
energy of the magnetic flux will remain true even if the higher Landau
levels are included.
(2) We solve the equations of motion for small
magnetic fields and show that there exists a solution with expectation
values for the tachyon such that the field strength vanishes.

\vspace{2ex}
\noindent
\emph{ (i) The tachyon potential}
\vspace{2ex}

We will show that 
tachyon wavefunctions with a Gaussian profile in $k_1$ cancel
a single unit of background flux.
For the
off diagonal fluctuation $W^+$, the tachyon wave function is given by 
\bea{profile1}
W^+(x^1, x^2) &=& 
\left(\frac{fQ}{4\pi l_1^2} \right)^{1/4} 
\int \frac{dk_1l_1}{2\pi}
e^{ -\frac{k_1^2}{2fQ}} W^+(k_1), \\ \nonumber
&=& 
\left(\frac{fQ}{4\pi l_1^2} \right)^{1/4} \frac{l_1}{2\pi} \sqrt{\pi
fQ}
e^{-\frac{fQ}{4}\left( (x^1)^2 + (x^2)^2 - 2i x^1x^2 \right) }
\left(
\begin{array}{c}
1 \\ i
\end{array}
\right).
\eea
Similarly the tachyon from the off diagonal fluctuation $W^-$  
with a  Gaussian profile is given by
\be{profile2}
W^-(x^1, x^2) 
= 
\left(\frac{fQ}{4\pi l_1^2} \right)^{1/4} \frac{l_1}{2\pi} \sqrt{\pi
fQ}
e^{-\frac{fQ}{4}\left( (x^1)^2 + (x^2)^2 + 2i x^1x^2 \right) }
\left(
\begin{array}{c}
1 \\ - i
\end{array}
\right).
\ee
We allow an arbitrary profile $\phi(k)$ for the diagonal fluctuations, 
the expansion of the gauge field with these fluctuations is given by
\bea{fluct}
A_a &=& 
A_a^{(0)} + 
\left(\frac{\pi l_2}{f Q l_1}\right)^{1/4}
\left(\chi  W^+_a L_+  +  \chi^* W^-_a L_- \right)  \\ \nonumber
&+& \frac{l_1l_2}{4\pi^2}
\int d^2k \phi(k) \frac{i\epsilon_{ab}k_b}{\sqrt{l_1l_2} |k|}
e^{ik\cdot x} L_3, 
\eea
where $\chi$ stands for the expectation value of the tachyon and 
we have
normalized $\chi$ for convenience. $A_a^{(0)}$ stands for
background gauge field in \eq{background}, 
for a single unit of flux we
have
\be{flux}
\int F_{12} dx^1 \wedge dx^2 = 2\pi
\ee
Therefore the field strength is given by $fQ = 2\pi /(l_1l_2)$.
Substituting the expansion \eq{fluct} 
in the Yang-Mills action \eq{YMaction} and integrating
out the diagonal fluctuations we obtain the following effective
potential for the tachyon.
\be{effact}
S= \frac{1}{4g_{YM}^2}\int dx^0 dx^3 
\left[\frac{16\pi^2}{l_1l_2}  + 
16 \left( -\frac{2\pi}{ \sqrt{l_1l_2}} |\chi|^2 + |\chi|^4 \right) 
\right].
\ee
We have provided the details of the evaluation of this potential in
appendix A. 
Note that the action is a perfect square and 
that the minimum of the
tachyon potential cancels the energy due to the background flux.
Therefore 
a Gaussian profile for the tachyon cancels a single unit of flux
and the tachyon drives the system to the vacuum.  
Though the wave function is a Gaussian it has support over the whole
range of the box, this is the reason that the constant flux density  
is canceled by the tachyon profile.
It will be interesting to find out the tachyon profile 
for larger units of flux

Now let us consider including the higher Landau levels in the
analysis. The expectation values for the diagonal fluctuations
$\phi(k)$ induces higher Landau levels to acquire expectation values.
Integrating the higher Landau levels  
will introduce higher order terms in the effective action.
Then the best way to analyze the effective potential of the
tachyon is to look at the complete action given by 
\be{squares}
S = -\frac{1}{2g_{\rm{YM}^2}} 
\int dx^1 dx^2 \rm{Tr} (F_{12}F^{12})
\ee
This is the sum of squares, therefore the minimum is obtained at
$F_{12} =0$. Since there is no conserved charge 
as $\int dx^1dx^2 \rm{Tr} (F_{12}) =0$,  there
is no obstruction for the field strength to vanish. 

\vspace{2ex}
\noindent
\emph{(ii) Solution of equations of motion for weak fields}
\vspace{2ex}

The equation of motion for the Yang-Mills action is given by 
$D^\mu F_{\mu\nu} = 0$. 
Consider the following expansion around the background \eq{rad-back}
\be{exp-rad}
A_a^3 = \frac{Qf}{2} \epsilon_{ab} x^b + \phi_a(x^1, x^2), 
\quad A_a^\pm = W_a^\pm(x^1, x^2),
\ee
Let the diagonal components, including the background field $f$ 
be of the $O(\epsilon^2)$ and the 
off diagonal components be of the  $O(\epsilon)$.
Let $\phi_a$ and $W_a^\pm$ satisfy the background gauge condition
\eq{back-gauge}. 
We expand $W_a^\pm$ in eigen functions of the operator
$O_{ab}^\pm = D^cD_c\delta_{ab} \mp 2i\epsilon_{ab}fQ$.
The equation of motion in terms of $\phi_a$ and $W_{ab}^\pm$ 
to $O(\epsilon^3)$ are given by
\bea{gauge-eom}
\partial_b 
F^3_{ba} = 0 \quad\quad\quad\quad\quad\quad&\;&
\\ \nonumber
\left( O_{ab}^+ W_{b}^+ -i W_b^+\partial_b\phi_a + i
\phi_b\partial_b W_a^+  
- i W_b^+(\partial_b \phi_a -\partial_a\phi_b)
\right. 
&\;&\\ \nonumber
\left. 
+  2 W_b^+(W_b^+W_a^- - W_a^-W_b^+)\right) &=&0 \\ \nonumber
\left( O_{ab}^- W_{b}^- +i W_b^-\partial_b\phi_a - i
\phi_b\partial_b W_a^-  
+ i W_b^-(\partial_b \phi_a -\partial_a\phi_b)
\right. 
&\;& \\ \nonumber
\left. 
+  2 W_b^-(W_b^-W_a^+ - W_a^+W_b^-) \right)&=&0
\eea
Here 
$F^3_{ab} = -Qf\epsilon_{ab} 
+ \partial_{a} \phi_{b} - \partial_{a} \phi_{b} +
2i(W_a^+ W_b^- - W_a^-W_b^+)$. We have  
also used the property that $F^\pm_{12} \sim
O(\epsilon^3)$ in obtaining the above equations.

We now solve these equations and show that the solution corresponds to
vanishing field strength.
The eigenfunctions of the operator $O_{ab}^\pm$ 
always have a Gaussian factor  with a width
$1/\sqrt{f}$, therefore derivatives of $W_i$ are of order
$O(\epsilon^3)$ and can be set to zero in the above set of equations. 
Then the equation of motion for $\phi_i$ is solved to $O(\epsilon^3)$ by
 $\phi_i=0$.
This makes it 
consistent to set coefficients of all the higher level Landau
wave functions to zero 
and work only with the tachyon. 
The off-diagonal components obey the following relations
\be{rel-llg}
W_1^+ = iW_2^+, \quad W_i^{-} = (W_i^{+})^* 
\ee
Using these relations the equation of motion for $W_1^+$ reduces to
\be{tac-eom}
-Qf W_1 + 4W_1 |W_1|^2 =0
\ee
Thus the solution is given by $|W_1^+|^2  = Qf/4$. 
Note that for this expectation value 
the field strength $F_{ij}^3$ and $F_{ij}^\pm$ vanish. 
Furthermore, this expectation value is
consistent with the fact the $W_i^\pm \sim O(\epsilon)$.
Thus the tachyon condenses so as to cancel the background flux
\footnote{
The fact that the expectation value of the tachyon comes out to be
constant is consistent with the fact that it is a Gaussian since the
Gaussian has width $1/\sqrt{f}$ and it is almost
constant for a large region.}.

\section{Constant magnetic fields in string theory}

In this section we present a brief review of 
exact string backgrounds representing 
constant magnetic fields in string theory. These form a special class
of plane wave solutions  found in \cite{Horowitz:1995rf}. 
There are embeddings of these plane wave solutions in heterotic string
theory which have localized tachyons. 
For weak magnetic 
fields there exists a tachyon which can be identified with
the Nielsen-Olesen tachyon. 
We show that this tachyon can be studied within supergravity if the
magnetic field arises from the $SO(32)$ gauge group of the heterotic
string.

\subsection{The bosonic string}

Though the bosonic string has
the usual bulk tachyon there are additional 
localized tachyons analogous to the Nielsen-Olesen tachyon 
in presence of magnetic fields. 
We will discuss the situation in detail for the case of the 
bosonic string
and we will be brief for the super string theories, 
a detailed discussion can be found
in \cite{Russo:1995cv, Russo:1995aj}. 

We first introduce their supergravity
description. 
Consider the following background in the bosonic string
\bea{back-bos}
ds^2 &=& dudv + a_i dx^i du + dx^i dx_i + dx^m dx_m,  \\ \nonumber
B_{iu} &=& a_i, \quad e^{2\Phi} = g_s^2, \\ \nonumber
\hbox{ with } 
a_i &=& \epsilon_{ij}  \frac{\tilde{f} }{2} x^j, \quad \hbox{ and } 
\tilde{f} =
\sqrt{\frac{\alpha'}{2}} f
\eea
Here $i,j \in \{ 1,2\} $, $m, n \in \{ 3, 4, \ldots 24\}$ and  
$u = \phi -t, v = \phi+t$. $t$ is the time coordinate and
$\phi$ is the last
spatial coordinate, it is compactified on a circle of radius $R$. 
$B_{iu}$ refers to the Neveu-Schwarz $B$ field, the dilaton is constant.
The solution in \eq{back-bos} has a null killing vector
$\frac{\partial}{\partial v}$ and is a special class of the 
plane wave solutions found in \cite{Horowitz:1995rf}. 
On Kaluza-Klein reduction of this plane wave
solution along $\phi$ 
one obtains a uniform magnetic field background in string theory.
Using the Kaluza-Klein anstaz on \eq{back-bos} we obtain
\bea{kk-bos}
ds^2 &=& -(dt + a_i dx^i)^2 + dx^idx_i + dx^m dx_m  \\ \nonumber
A_i^{(1)} = \frac{\alpha'}{R^2}a_i &\quad& A_i^{(2)} = -
\frac{\alpha'}{R^2} a_i \\ \nonumber
B_{ti} = a_{i}, &\quad& e^{2\Phi} = \frac{g_s^2\sqrt{\alpha'}}{R}
\eea
Here we have labelled the $U(1)$ 
gauge fields according to the convention
given in \cite{Sen:1994fa}. Now it is easy to see that the gauge
potentials correspond to constant magnetic fields. 
This solution is different from 
the Melvin background
\cite{Melvin:1964qx} 
which also contains magnetic fields.
The magnetic fields in the Melvin universe
is non-uniform,  
the magnetic field strength decreases from a finite value at the
origin to zero at infinity.

The world sheet action on the background given in \eq{back-bos} 
reduces to
\bea{ws-bos}
S= -\frac{1}{4\pi\alpha'} \int d\sigma^+ d\sigma^-
&\left( \partial_+ u\partial_- v +\partial_+ v\partial_- u +
\partial_+ X \partial_- \bar{X} +\partial_+ \bar{X} \partial_- X
 \right. \\ \nonumber 
&\left. -i\tilde{f} (
X \partial_+ \bar{X} - \bar{X} \partial_+ X ) \partial_- u
+ 2 \partial_+ x^m \partial_- x_m
\right)
\eea
Here $\sigma^+ = \sigma^0+ \sigma^1$ and $\sigma^- = -\sigma^0 +
\sigma^1$.
Note that the interaction terms proportional to $\tilde{f}$ 
is chiral. 
From this world sheet action it is easy to see that the plane wave
representing a constant magnetic field background is exact to all
orders in $\alpha'$.
Consider the following redefinitions
\be{redef}
X= e^{i\tilde{f}\tilde{u}(\sigma^-) } Y , \quad \bar{X} = 
e^{-i\tilde{f}\tilde{u}(\sigma^-)}\bar{Y}, 
\ee
where $\tilde{u}(\sigma^-)$ refers to the right moving part of $u$.
Then the action in \eq{ws-bos} reduces to a free action in $u, v, Y,
\bar{Y}, x^m$. We can quantize the world sheet action in \eq{ws-bos}
by fixing the light cone gauge so that
\be{light-cone}
u = u_0 + p\sigma^+ - \tilde{p} \sigma^-
\ee
Then as the coordinates $X$ and $\bar{X}$ are single valued, 
the free coordinates $Y, \bar{Y}$ obey twisted boundary
conditions.
\be{bc-y}
Y(\sigma^0 , \sigma^1 + 2\pi) = e^{2\pi i\tilde{f}\tilde{p}} Y(\sigma^0,
\sigma^1),  \quad 
\bar{Y} (\sigma^0, \sigma^1 + 2\pi) = e^{-2\pi i\tilde{f}\tilde{p}} 
\bar{Y} (\sigma^0, \sigma^1)
\ee
With these boundary conditions these fields can be mode expanded as
\bea{bos-mode}
Y &=& i\sqrt{\alpha'} 
\sum_{n=-\infty}^{\infty} \left( \frac{|n-\nu|^{1/2} }{n-\nu} a_n
e^{-i(n-\nu)\sigma^+ }
+ \frac{| n+\nu|^{1/2} }{n+ \nu} \tilde{a}_n  e^{i(n+\nu)\sigma^-}
\right), \\ \nonumber
\bar{Y} &=& i \sqrt{\alpha'} \sum_{n= -\infty}^{\infty} \left( 
\frac{ |n+\nu|^{1/2} }{n+ \nu}  b_n e^{-i(n+\nu) \sigma^+} +
\frac{ | n-\nu|^{1/2} }{ n-\nu} 
\tilde{b}_n e^{i(n-\nu) \sigma^-}\right),
\eea
here $\nu = \tilde{f}\tilde{p}$ and with out loss of generality we can
choose $0<\nu <1$. The commutation relations for the left moving 
oscillators are
\bea{bos-comm}
[a_n , b_{-m}] = \delta_{n,m},  &\quad& 
[b_n, a_{-m}] = \delta_{n,m},  \quad
 \hbox{ for } n, m  = 1, 2, \ldots   \\ \nonumber
\hbox{ and }[b_0, a_0] &=& 1,  
\eea
with similar commutation relations for the right movers.
From these mode expansions and commutation relations
it is easy to write down the spectrum of
bosonic strings in the background \eq{back-bos}.
The spectrum for $k^m=0$ is given by
\bea{bos-spec}
M^2 &=&  \left(\frac{n}{R}\right)^2 + \left(
\frac{wR}{\alpha'} \right)^2 
+\frac{2}{\alpha'} \left( N+ \tilde{N} - 2 + \nu(1-\nu)  \right)
\\ \nonumber
&+& 2(\tilde{M}+Q_R)f(l+ \frac{1}{2} ) - 2(\tilde{M}+Q_R)fS  
\\ \nonumber
\hbox{ with } \tilde{N} - N &=& nw
\eea
where $n$ and $w$ are the momentum and winding numbers along the
compact direction $\phi$,  $\tilde{M}$ and 
$Q$, the right moving charge are given by
\be{left-charge}
\tilde{M} = \sqrt{\frac{\alpha'}{2}} M, \quad
Q_R= \sqrt{\frac{\alpha'}{2}} \left( \frac{n}{R} - \frac{w R}{\alpha'}
\right)
\ee
$l$ stands for the occupation number 
of the zero mode left oscillators given by $a_0b_0$, 
this is the Landau level. $S$ is the
angular momentum of the left moving oscillators in the $Y, \bar{Y}$
plane, for the bosonic string it is given by
\be{bos-ang}
S  
= \sum_{n=1}^{\infty} \left( b_{-n} a_{n} - a_{-n} b_{n} \right)
\ee
The mass formula in \eq{bos-spec} is
valid for $Q_R>0$, the same formula with $Q_R\rightarrow -Q_R$ and
$S\rightarrow -S$ is true for $Q_R<0$. 

We now show that there is a tachyon which corresponds to the
Nielsen-Olesen tachyon in the bosonic string.
Consider the state with $n=1, w = -1, N= 1, \tilde{N}=0 $ 
at the self dual radius $R=\sqrt{\alpha'}$, 
then $S= \pm 1$ and the (mass)$^2$ of this
state reduces to
\be{no-tac}
M^2 = 2(\tilde{M} + Q)f(l + \frac{1}{2} ) \mp 2(\tilde{M} + Q) fS +
\frac{2}{\alpha'}  \nu (1-\nu) 
\ee
where $Q= \sqrt{2}$. 
Ignoring the zero point energy we 
see that for weak magnetic field $f\ll 4Q/{\alpha'}$ 
this state has the same
mass formula as the Nielsen-Olesen tachyon in \eq{energy}.
In general when the compact direction is not at the self dual point
the off diagonal gauge bosons charged under the background $U(1)$
gauge field are massive.  
In fact the contribution to the
(mass)$^2$ in \eq{bos-spec} from the winding and momentum modes are
the masses of the off diagonal bosons.
Note that the states which are tachyonic in general 
have both winding and
momentum quantum numbers and higher spins, 
therefore they cannot be seen in
supergravity. 
In superstring theories with constant magnetic field the (mass)$^2$
formula basically remains the same with minor modifications.
Superstring theories do not have the zero point energy making the
identification with the Nielsen-Olesen tachyon more precise.

\subsection{Type II A/B string}

The magnetic flux background in \eq{back-bos} can be embedded in
either type IIA or type IIB string theory. 
The world sheet action can 
be obtained from the bosonic action
in \eq{ws-bos} by promoting the bosonic fields to super fields, 
it is given by
\bea{ws-typeii}
S&=&  S_{\rm{Bosonic}} +  S_{\rm{Left}} + 
S_{\rm{Right}} \\ \nonumber
S_{\rm{Left}} &=& 
-\frac{1}{4\pi \alpha'} \int d\sigma^+d\sigma^-
\left( i\partial_- \psi^v \psi^u + i \partial_- \psi^u \psi^v 
+ i \partial_-\psi^X \psi^{\bar{X}} + i \partial_- \psi^{\bar{X}}
\psi^{X}\right)  \\ \nonumber
& & \quad\quad\quad+ 2 i \partial_- \psi^m \psi_m  
 + \tilde{f} \partial_- u \left(
\psi^X\psi^{\bar{X} } - \psi^{\bar{X} } \psi^X \right)
 + \tilde{f} \partial_-\psi^u \left( X\psi^{\bar{X}} - \bar{X} \psi^X
 \right)  \\ \nonumber
 S_{\rm{Right}} &=& - \frac{1}{4\pi\alpha'}
 \int d \sigma^+ d\sigma^- 
 \left( i \tilde{\psi}^v \partial_+ \tilde{\psi}^u + i \tilde{\psi}^u
 \partial_+ \tilde{\psi}^v  
+ i\tilde{\psi}^X\partial_+ \tilde{\psi}^{\bar{X}} 
+ i \tilde{\psi}^{\bar{X}} \partial_+ \tilde{\psi}^X \right)\\ \nonumber
& & \quad\quad\quad +2 i \tilde{\psi}^m \partial_+ \tilde{\psi}_m 
 + \tilde{f} \tilde{\psi}^u \left(
 X\partial_+ \tilde{\psi}^{\bar{X}} - \bar{X} \partial_+
 \tilde{\psi}^X + \tilde{\psi}^X \partial_+ \bar{X} -
 \tilde{\psi}^{\bar{X}} \partial_+ X \right) 
 \eea
Here the superscripts on the fermions label their super partners, 
$S_{\rm{Bosonic}}$ stands for the bosonic action in \eq{ws-bos} and
now $m$ runs from  $3, 4 \ldots 8$. The compact direction $\phi$ is the
$9$th coordinate.
Again this action is conformal to all orders in $\alpha'$ and it can
be quantized in the light cone gauge. 

The uniform magnetic field background in  IIA/B string theory
preserves 16 of the 32 supersymmetries
\cite{Horowitz:1995rf,Bergshoeff:1993cw}.
The Nielsen-Olesen type tachyons present in the 
bosonic string are projected out here by the GSO projection. 
As an example consider the same state 
we discussed for the case of the bosonic string with the quantum
numbers $ n=1, w= -1, N=1, \tilde{N}=0$
in the Neveu-Schwarz sector at the self dual radius, this
state corresponds to off diagonal gauge bosons. 
As this state has world sheet 
fermion number $-1$ for the left movers it is projected
out by the GSO projection.

\subsection{Heterotic string}

The most symmetric way to embed the magnetic flux background in
the heterotic string would be such that the action 
$S_{\rm Left}$ in \eq{ws-typeii} came from
the left moving fermions and the $S_{\rm Right}$ from the fermions
corresponding to the $SO(32)$ gauge bosons. However the signature on
the right moving fermions is Lorentzian while the 
world sheet fermions
corresponding to the gauge bosons in the heterotic string have
Euclidean signature. Therefore this totally symmetric 
embedding is prohibited. 
We discuss the three ways the magnetic flux background
\eq{back-bos} can be embedded in heterotic string theory. 
These are best described from the world sheet point of view.

\vspace{2ex}
\noindent
(i)\emph{Left Truncation}
\vspace{2ex}

Let the fermions corresponding to the gauge bosons be left movers.
Then the world sheet action for the left truncated embedding is
obtained by dropping the action $S_{\rm{Left}}$. The world sheet
action is given by
\be{left-trunc}
S = S_{\rm{Bosonic}} + S_{\rm{Right}} -\frac{1}{2\pi \alpha'} \int 
d\sigma^+ d\sigma^- i \partial_- \lambda^A \lambda_A
\ee
Here $\lambda^A$ with $A= 1, \ldots 32$ are the left moving fermions
corresponding to the $SO(32)$ gauge bosons.
This embedding of the uniform magnetic field in heterotic string
theory preserves $1/2$ of the supersymmetries of the heterotic string
\cite{Horowitz:1995rf}. In light cone gauge we can 
set $\tilde{\psi}^u=0$.
Therefore all terms proportional to the interaction $\tilde{f}$ drops
out in $S_{\rm{Right}}$. Then the 
quantization of the fermions is trivial.
The off diagonal gauge bosons 
at the self dual radius which are charged under the background
$U(1)$ and which would have been the Nielsen-Olesen tachyons 
are projected out by the GSO projection. In this case the gauge bosons
 from the $SO(32)$ gauge group is not charged under the background
$U(1)$ therefore they are also not tachyonic.

\vspace{2ex}
\noindent
(ii)\emph{Right Truncation}
\vspace{2ex}

Now consider the fermions corresponding to the gauge bosons be right
movers. The right truncated model is obtained by dropping the action
$S_{\rm{Right}}$. The world sheet action is given by
\be{right-trunc}
S= S_{\rm{Bosonic}} + S_{\rm{Left}} -\frac{1}{2\pi\alpha'} \int
d\sigma^+ d\sigma^- i \tilde{\lambda}^A \partial_+\tilde{\lambda}_A
\ee
This embedding of the uniform magnetic field in heterotic string is
non-supersymmetric and there are tachyons. 
The world sheet action can be quantized in the light cone gauge, the 
mass spectrum in the Neveu-Schwarz sector is given by
\cite{Russo:1995aj}
\be{het-spec}
M^2 = Q_L^2 + 2f(Q_R + \tilde{M})(l + \frac{1}{2}) - 2f(Q_R +
\tilde{M}) S_L + \frac{4}{\alpha'} ( N_L -\frac{1}{2} )
\ee
Here we have eliminated the level number of the right movers using the
level matching condition. $Q_L$ is the left moving charge given by
\be{het-def}
Q_L = \sqrt{\frac{\alpha'}{2}} \left( \frac{n}{R} + \frac{wR}{\alpha'}
\right), 
\ee
the left moving angular momentum $S_L$ in the $Y, \bar{Y}$ plane 
has contribution both from the bosonic oscillators and the fermionic
ones. It is easy to see that this model has Nielsen-Olesen type
instabilities. Consider the states with 
$N_L = 1/2, Q_R = \sqrt{2}, Q_L=0$ at the self dual radius where the
gauge group is enhanced to $SU(2)$, 
the (mass)$^2$ of this state is
given by
\be{het-mass}
M^2 = 2f(Q_R + \tilde{M})(l + \frac{1}{2})  \mp 2f( Q_R + \tilde{M})
\ee
For small values of $f$ this mass spectrum 
reduces to the  spectrum of fluctuations of the off-diagonal gauge
bosons  in a $SO(3)$ gauge theory with a uniform magnetic field.
Therefore there is a Nielsen-Olesen instability of small $f$. 
In fact there is an infinite tower of tachyons from higher levels which
are charged with respect to the background $U(1)$. Consider the states
with $N_L = 1/2 + m^2, Q_R= m\sqrt{2}, Q_L=0, l =0$ 
at the self dual radius. For large $m$,  $S_L\sim m^2$, 
these states are tachyons for $m > 2\sqrt{2}/(\alpha' f)$. 
Thus for arbitary small magnetic
fields there are tachyons at higher levels with high spins
\cite{Russo:1995aj}.
All the tachyons have both winding modes and momentum modes or higher
spins
therefore they cannot be studied in supergravity. If on the other hand
there exists an action for the enhanced gauge group coupled with
gravity the tachyon in the lowest level, $N_L = 1/2$ 
can be studied. This is conveniently achieved by
embedding the $U(1)$ magnetic field in the $SO(32)$ gauge group of the
heterotic string.

\vspace{2ex}
\noindent
(iii)\emph{ Magnetic field from the $SO(32)$ gauge group}
\vspace{2ex}

Consider the Kaluza-Klein compactification of the 
right truncated model at the self dual radius. 
The supergravity solution is given in \eq{kk-bos} with
$R=\sqrt{\alpha'}$. 
We can obtain the supergravity solution with the magnetic field in the
internal gauge group by applying a $SO(1,17)$ T-duality transformation.
This T-duality transforms the $U(1)$ gauge fields arising from the
Kaluza-Klein direction to a gauge field arising from the $SO(32)$
gauge group, it leaves the metric and the $B$ field invariant.
We use the notation and formulae in \cite{Sen:1994fa}. 
At the self dual radius the moduli
matrix $M$ is a $18\times 18$ identity matrix. 
The $SO(1,17)$ transformation is given by
\be{t-dual-trans}
\Omega = 
\left(
\begin{array}{cccc}
\frac{1}{2} &  \frac{1}{2} & \frac{1}{\sqrt{2}} & 0\\
\frac{1}{2} &  \frac{1}{2} & - \frac{1}{\sqrt{2}} &0 \\
\frac{1}{\sqrt{2}} & -\frac{1}{\sqrt{2}} &0 & 0 \\
0 & 0 & 0 & I_{15}
\end{array}
\right)
\ee
Note that $\Omega$ satisfies $\Omega^{T} L \Omega =L$ 
with $L$ given by
\be{def-met}
L= \left(
\begin{array}{ccc}
0 & 1 & 0 \\
1 & 0 & 0 \\
0 & 0& -I_{16}
\end{array}
\right)
\ee
here $I_{15}, I_{16}$ are $15\times 15$ and $16\times 16$ dimensional
identity matrices respectively. This $SO(1,17)$ 
transformation basically exchanges
the the right moving $S^1$ with a $S^1$ from the right moving 
internal torus.
The gauge fields transform as $V_i \rightarrow \Omega V_i$ where
$V_i$ is a $18$ dimensional vector.
For the solution in \eq{kk-bos} $V_i$ is given by
\be{het-gauge}
V_i = 
\left(
\begin{array}{c}
A_i^{(1)}\\
A_i^{(2)}\\
0 \\
\vdots\\
0
\end{array}
\right)
\ee
$\Omega V_i$ has only one $U(1)$ gauge field embedded in the $SO(32)$
gauge group, given by $A_i^{\rm{Int}} = \sqrt{2} a_i$. 
The world sheet theory with the gauge field in the $SO(32)$ gauge
can also be quantized in the light cone gauge 
\cite{Russo:1995aj}. 
T-duality at the self dual radius is a gauge symmetry, therefore one
can think of this T-duality also as  a gauge transformation in the
enhanced gauge group \footnote{The author thanks Ashoke Sen for
pointing this out.}.
At the self dual point the gauge group is $SU(2)_R
\times SO(32)$. 
The gauge field in the background \eq{kk-bos} is
embedded in the $U(1)$ of $SU(2)_R$. The  
background gauge field is given by 
\be{int-gauge}
A_i^{(R)} = \frac{1}{\sqrt{2}} (A_i^{(1)} - A_i^{(2)}) = \sqrt{2}a_i
\ee
Now the gauge transformation which exchanges the $SU(2)_R$ with an
$SU(2)$ of $SO(32)$ embeds the background gauge field in the $SO(32)$
gauge group.

Since the background with magnetic field in the $SO(32)$ group
is obtained by a
T-duality from the right truncated model, the spectrum is the same as
in \eq{het-spec} with $Q_L =0$\footnote{We have decompactified the
original $S^1$ we started out with.}. 
and $Q_R$ being the momentum on the even 
self-dual
lattice of the $SO(32)$ gauge group.
For convenience we write down the spectrum 
\be{het-int}
M^2 =  2f(Q_R + \tilde{M}) (l + \frac{1}{2}) - 2f(Q_R + \tilde{M}) S_L +
\frac{4}{\alpha'} (N_L- \frac{1}{2})
\ee
As in the discussion of the right truncated model, states with $N_L=
1/2, Q_R= \sqrt{2}$ corresponds to off-diagonal fluctuations charged
under the the background $U(1)$. But unlike that case it is easy to
obtain these states as fluctuations around the supergravity
background as the low energy effective action contains
the full $SO(32)$ gauge fields including the off-diagonal
fluctuations. There are tachyons with higher spins as in the right
truncated model and these tachyons cannot be seen in supergravity.
We will call this lowest lying tachyon as the Nielsen-Olesen tachyon.

\section{The Nielsen-Olesen instability in supergravity}

We have seen that 
to study the Nielsen-Olesen tachyon in supergravity we need to 
embed the uniform magnetic field in the $SO(32)$ gauge group. 
In this section we discuss some important properties of this
background and perform 
a linearized fluctuation analysis around the
supergravity background and determine the tachyon explicitly.

To set up notations and conventions we write down the bosonic part of
the low energy effective action of the heterotic string theory is
given by
\be{hetaction}
S= \frac{1}{(2\pi)^7 (\alpha')^4 }  \int d^{10} x
\sqrt{-g} e^{-2\Phi} \left[ R + 4 \partial_\mu \Phi
\partial^\mu \Phi 
-\frac{1}{8} \frac{\alpha'}{2}  
 \rm{Tr} ( F_{\mu\nu} F^{\mu\nu} ) -\frac{1}{12}
 H_{\mu\nu\rho}H^{\mu\nu\rho} \right].
\ee
where $R$ is the Ricci scalar, $F_{\mu\nu}$ denotes the non-Abelian
field strength. The gauge group is $SO(32)$ with the gauge fields in
the vector representation of $SO(32)$. 
There is an extra factor of $1/2$ in the action for the gauge field in
the above equation as we have normalized the
generators so that $\rm{Tr}(T^a T^b) = 2\delta^{ab}$.
$H_{\mu\nu\rho}$ is the field
strength associated with the $B_{\mu\nu}$ field
\bea{hfield}
H_{\mu\nu\rho} &=& \partial_\mu B_{\nu\rho} -\frac{1}{4}
\frac{\alpha'}{2}\rm{Tr} \left(
A_\mu F_{\nu\rho} - \frac{i}{3} A_\mu[A_\nu, A_\rho] \right) \\
\nonumber
&+& \hbox{cyclic permutations of} \quad \mu, \nu, \rho. 
\eea
Performing the T-duality in \eq{t-dual-trans} on the solution in
\eq{kk-bos} we obtain 
\bea{maghet1}
ds^2 &=& -(dt + a_i dx^i)^2 + dx^idx_i + dx^mdx_m \\ \nonumber
A_i = \sqrt{2} \frac{\epsilon_{ij} fx^j}{2}, &\quad& e^{2\Phi} =
g_s^2, \quad B_{ti} = a_i
\eea
where we have scaled the $U(1)$ gauge field by $\sqrt{2/\alpha'}$ so
that it has dimensions of mass. It is more convenient to perform
calculations in the Landau gauge. 
We can perform the following gauge and co-ordinate transformation to 
convert the background to the Landau gauge
\bea{g-trans}
t\rightarrow t+ \frac{\tilde{f}}{2} x^1x^2,   \\ \nonumber
A_i \rightarrow A_i + \partial_i \alpha, \quad \alpha =
\frac{f}{2}x^1x^2.
\eea
After the gauge transformation we obtain the following solution in the
Landau gauge.
\bea{maghet}
ds^2 &=& -(dt+ a dx^1)^2  + dx^idx_i  + dx^m dx_m\\ \nonumber
a = \sqrt{\frac{\alpha'}{2} } fx^2, &\quad&
A_1 = Qfx^2 L_3 \\ \nonumber
B_{t1} =\tilde{f} x^2, &\quad& e^{2\Phi}  = g_s
\eea
The above metric is not asymptotically flat, in fact it  has a
constant curvature given by
\be{curv}
R= \frac{1}{2}\tilde{f}^2, \quad R_{\mu\nu}R^{\mu\nu} = \frac{3}{4}
\tilde{f}^4
\ee
Therefore for small field strength $\tilde{f}$ we can trust our
analysis in supergravity and as the dilaton is constant, 
we can supress string loop
corrections by tuning $g_s$ to be small.

\subsection{Fluctuations around a constant magnetic field 
in supergravity}

In this section we determine how the Nielsen-Olesen tachyon occurs
as a fluctuation of the fields in supergravity, we will explicitly show
that the spectrum of fluctuation is given by \eq{het-mass}.
We have seen that the Nielsen-Olesen tachyon is an off diagonal
fluctuation of the gauge field. Therefore consider the following
fluctuations of the solution in \eq{maghet} which involves only off
diagonal components of the gauge field. 
\bea{back-fluct}
A_\mu = \delta_{1\mu} fx^2L_3 + W^+_\mu L_+ + W^-_\mu L_- \\ \nonumber
g_{\mu\nu} = g_{\mu\nu}^{(0)}, \quad B_{\mu\nu} = B_{\mu\nu}^{(0)},  
\quad \Phi = \Phi^{(0)} 
\eea
here the superscript $^{(0)}$ refers to the background in \eq{maghet}.
The equation of motion for these fluctuations reduce to
\be{cov-eom}
\hat{D}^\nu\hat{D}_\nu  W_\mu - 2i [F^{\nu}_{\;\;\mu},W_\nu] 
+ R_{\mu}^{\;\;\nu}W_\nu  
=0
\ee
here $W_\mu$ stands for the off diagonal fluctuations $W_\mu^+$ and
$W_\mu^-$ and
$\hat{D}_\mu$ refers to the covariant derivative given by
$
\hat{D}_\mu W_\nu= \partial_\mu W_\nu
-\Gamma_{\mu\nu}^\rho W_\rho + i[A_\mu, W_\nu]
$.
In \eq{cov-eom} we have imposed the background gauge condition 
\be{back-gauge-g}
\hat{D}^\mu W_\mu =0
\ee

To solve the equations  \eq{cov-eom} and \eq{back-gauge-g} 
for fluctuations it is convenient to use an
ansatz. We will demonstrate it for the tachyons in the lowest Landau
level.  Consider the following off diagonal fluctuations
\bea{t-fluct}
W^+_i &=& \left( \frac{f(Q+\tilde{M})}{4\pi l_1^2} \right)^{1/4} 
e^{-iMt} e^{-ik_1x^1} e^{-\frac{f}{2}(Q+\tilde{M})
\left(\frac{k_1}{f(Q+\tilde{M})} - x^2 \right)^2 }
\left(
\begin{array}{c}
1\\ i 
\end{array}
\right) \\ \nonumber
W^-_i &=& \left( \frac{f(Q+\tilde{M})}{4\pi l_1^2} \right)^{1/4} 
e^{iMt} e^{-ik_1x^1} e^{-\frac{f}{2}(Q+\tilde{M})
\left(\frac{k_1}{f(Q+\tilde{M})} + x^2 \right)^2 }
\left(
\begin{array}{c}
1\\ -i 
\end{array}
\right)  \\ \nonumber
\hbox{with} & &  W_0^\pm = W_m^\pm =0
\eea
Substituting the above ansatz in  \eq{cov-eom} and
\eq{back-gauge-g} one can show that the
equations of motion and the background gauge conditions 
are satisfied if
\be{t-energy}
M^2 = -f(Q+ \tilde{M})
\ee
This is the condition on the spectrum of for the tachyon in the 
lowest Landau level given in \eq{het-mass}. The excited state in the
$l=0$ Landau level is obtained by interchanging the two component
vectors in \eq{t-fluct}
\be{inter-change}
\left(
\begin{array}{c}
1\\ i 
\end{array}
\right)
\leftrightarrow
\left(
\begin{array}{c}
1\\ -i 
\end{array}
\right)
\ee
Higher Landau levels are obtained by replacing the Gaussian in
\eq{t-fluct} by the appropriate wave function of the harmonic
oscillator. Thus the spectrum given in \eq{het-mass} is reproduced.
Note that the tachyons are localized and
for small values of the field strength $f$ the spectrum reduces to
that of that of off diagonal fluctuations in the $SO(3)$ 
field theory found by Nielsen-Olesen. 
For later use we remark that the wave functions of the type 
constructed here with $M=0$ are eigen functions of the operator
\be{operator}
O^{\pm}_{ij}= 
\hat{D}^k \hat{D}_k \delta_{ij} \mp f\epsilon_{ij} + R_{ij},
\ee
with eigen values proportional to $f$.

\section{Condensation of the Nielsen-Olesen tachyon in supergravity}

In this section we argue that the Nielsen-Olesen tachyon in
supergravity drives the background \eq{maghet} to flat space.
Our argument rests on the three evidences discussed below.

\subsection{The S-dual in type I theory}

We have seen in the previous section that when the uniform magnetic
field is embedded in the $SO(32)$ gauge group of the heterotic string
the off diagonal fluctuations which are charged with respect to the
background $U(1)$ contain tachyonic modes. In the heterotic string
these are localized closed string modes in the twisted sector.
S-duality of the heterotic string maps it to type I string. 
From the mass formula in \eq{het-int} we see that these modes are
charged under the background $U(1)$ in heterotic string, hence one
would expect these modes to become open string modes in the type I
string. Therefore, it is possible to derive a decoupling limit in
which the closed strings decouple. Then the tachyon condensation
process reduces to that of tachyon condensation process in
field theory discussed in section 2. The tachyon drives the system to
the Yang-Mills vacuum. Thus we would expect 
by continuity,  that turning on the type I coupling the tachyon will
drive the system to flat space.  We now discuss this argument in
detail. 

Consider the S-dual of the constant magnetic field background in
heterotic string with the magnetic field from the $SO(32)$ gauge group
given in \eq{maghet}. 
S-duality involves the field redefinitions given by.
\be{s-dual}
\phi_{(I)} = -\Phi_{(H)}, \quad 
ds^2_{(I)} = e^{-\Phi_H} ds^2_{(H)}, \quad  B_{ti}^{(I)} =
B_{ti}^{(H)},  
A_\mu^{(I)} = A_\mu^{(H)} 
\ee
Using this map we obtain the solution
\bea{mag-typei}
ds^2 &=& -(dt + b dx^1)^2  + dx^i dx_i + dx^m dx_m \\ \nonumber
b = 
\sqrt{ \frac {\alpha'}{2g_I } } f x^2, &\quad&
A_1 = Q\frac{1}{g_I} f x^2 L_3 \\ \nonumber
B_{ti}^{RR} =  
\sqrt{\frac{\alpha'}{2g_I}}  f x^2, &\quad&
e^{\Phi_I} = g_{I} = \frac{1}{g_s}
\eea
Here we have rescaled coordinates by $1/\sqrt{g_I}$
to get rid of the pre-factors in front of the metric. Under the duality
map given in \eq{s-dual} the RR 2-form $B^{RR}$ does not couple to the 
dilaton by the usual prefactor $e^{-2\Phi_I}$, to ensure that the RR
form couples in the usual way we have 
scale the  RR form by
$g_I$ so that it appears in the action with the usual dilaton
coupling. This coupling of the RR form 2-form is the natural one if
one derives the low energy effective 
action using scattering amplitudes in
string theory \cite{Polchinski:1998rr}. The 
S-duality has converted the gauge field to the
open string sector. 

It is easy to see from the above background that the
closed string fields is suppressed by a factor of $\sqrt{g_I}$ with
respect to that of the open string gauge field. We can make use of
this to derive the decoupling limit.
This is essentially a limit in which the
back reaction due to the presence of the magnetic flux which results
in a change in the closed string fields,
the metric and the Ramond-Ramond background is
neglected.  The limit is given by
\be{decoupling}
 g_I \rightarrow 0, \quad \hbox{ with } \quad \frac{f}{g_I} =  h
\quad \hbox{ held fixed}
\ee
Note that the decoupling  limit in type I 
corresponds to infinite coupling in the heterotic
side.
In this limit the background in \eq{mag-typei} reduces to a flat
metric with the Ramond-Ramond flux set to zero. The gauge field is
given by
\be{typei-gauge}
A_1 =  Qh x^2L_3
\ee
The type I open string in this constant gauge field can be quantized
\cite{Fradkin:1985qd, Abouelsaood:1987gd,Nesterenko:1989pz,
Bachas:1995ik}. 
From the spectrum it can be seen that there exists a
Nielsen-Olesen tachyon which survives in the $\alpha'\rightarrow 0$
limit. The tachyon condensation 
process  can now be studied in field theory and 
from the discussion in section 2 we can conclude that it drives the
system to the vacuum. 
On the heterotic side this implies that at strong coupling the tachyon
will drive the system to flat space. It is natural to expect 
by continuity that
lowering the coupling to weak coupling on the heterotic the result
would not change. There could be a possible phase transition which
might affect this conclusion therefore we will discuss the 
the tachyon condensation process directly in the heterotic theory in the
next subsections.

As an aside we remark that since the tachyons in the background
\eq{mag-typei} are in the open string sector, 
they can be studied using Berkovits' open super string field theory.

\subsection{The closed string tachyon potential}

In this section we obtain the closed string tachyon potential for the
background with uniform magnetic field in the $SO(32)$ group of the
heterotic string theory given in \eq{maghet1} in supergravity.
We show that the minimum of the tachyon potential is flat space.
To obtain the tachyon potential we have to define a notion of energy
for the space \eq{maghet1}, 
as this space is not asymptotically
flat it is difficult to define a notion of energy. 
However, the metric
in \eq{maghet1} admits a killing vector $\xi^\mu =(1, 0, 0)$, 
(we have listed only the relevant co-ordinates). This can be used to
show that the following integral of the stress energy tensor is a
conserved quantitiy
\be{def-ene-fucn}
E = 
\frac{1}{(2\pi)^7(\alpha')^4 g_s^2}
\int d^9 x \sqrt{h} T_{\mu\nu} \eta^\mu\xi^\nu, 
\ee
here $\eta^\mu$ is the unit normal to a  space like surface
$\Sigma_t$ of constant $t$ and $h_{ij}$ is the induced metric on 
$\Sigma_t$.
All fluctuations of the metric should preserve the constraint
equations of gravity, as these are constraints for intial velocities.
Therefore, to obtain the tachyon potential we evaluate the energy
functional 
given in \eq{def-ene-fucn} on the fluctuations which are 
subject to the gravity
constraints
\footnote{This method is inspired by discussions with M. Gutperle, M.
Headrick and S. Minwalla during  the collaboration 
\cite{David:2001vm}.}. 

We first restrict our fluctuations to preserve the form of the metric,
and the $B$-field of the background given by
\bea{fluctansatz}
ds^2 &=& -(dt - g_i(x^1, x^2) dx^i) ^2 + 
dx^i dx_i + dx^mdx_m \\ \nonumber
B_{ti} = b_i(x^1, x^2) &\quad&  e^{\Phi} = g_s \quad A_i =
A_{i}(x^1, x^2)
\eea
We have allowed dependence of the functions only on the $x^1, x^2$
directions as the zero momentum tachyon $(k^m=0)$, depends only on these
coordinates. Since the background and the tachyons do not involve the
dilaton we have assumed it to be constant. We also assume that the
background and the fluctuations are small, so that deviations from
flat space will be small. 
Now we use the 
constraints from the supergravity equations to eliminate the
metric, and the $B$ field in terms of the gauge field. 
The constraint is given by the $00$ component of the Einstein
equation. 
\be{gra-const}
R_{00} -\frac{1}{2} g_{00}R = \frac{1}{4} H_{0\mu\nu}
H_{0}^{\;\;\mu\nu} - 
\frac{1}{2}g_{00}\left( \frac{1}{12}H_{\mu\nu\rho}H^{\mu\nu\rho} +
\frac{\alpha'}{16} \rm{Tr} (F_{\mu\nu} F^{\mu\nu}) \right)
\ee
Here we have set the dilaton to be constant. 
The constraint can be satisfied if 
$g_i= b_i$ and by requiring \footnote{This is in fact the dilaton
equation of motion.}
\be{dil-constraint}
-(\partial_1 g_2 - \partial_2 g_1)^2 + \frac{\alpha'}{8}
\rm{Tr}(F_{12}
F_{12}) =0.
\ee
Thus 
the only degree of freedom left over is that of the gauge field. 
Using these constraints in the energy formula in \eq{def-ene-fucn} and
retaining only the leading order term in the fields we obtain
\footnote{For the background in \eq{fluctansatz} the Einstein frame is
the same as the string frame since the dilaton is constant.}
\bea{energy-func}
E &=& 
\frac{1}{(2\pi)^7(\alpha')^4 g_s^2}
\int d^{9} x
\left(R_{00} -\frac{g_{00}}{2} R\right)  \\ \nonumber
&=&
\frac{1}{8 (2\pi)^7(\alpha')^3 g_s^2}
\int d^{9}x \rm{Tr}(F_{12}F_{12})
\eea
Tachyon condensation should proceed by minimizing this energy.
The energy functional is proportional to the negative of the Yang-Mills
action. Thus
the calculation of the tachyon potential reduces to that
performed in section 2. 
Considering fluctuations involving only the tachyon and 
integrating out the diagonal fluctuation 
$\phi$ we obtain the following potential
\be{het-t-pot}
E =  
\frac{1}{8(2\pi)^7(\alpha')^4 g_s^2}
\int d^8x \left( \frac{4\pi}{\sqrt{l_1l_2}} - |\chi|^2 \right)^2
\ee
here we have chosen the background field strength $f = 2\pi/l_1l_2$, 
considering a  
single unit of flux on the $1-2$ plane and regulated the system in a
large box of size $l_1$ and $l_2$. 
The minimum of the tachyon potential cancels the
energy due to the background magnetic field. 
Though our analysis above was restricted only to fluctuations of the
gauge field in the lowest Landau level we have seen in section 2, that
even on the inclusion of the higher levels the potential is minimized
at $F_{12} =0$. From the constraint \eq{dil-constraint}
this implies that the curvature which is proportional 
$(\partial_1 g_2- \partial_2 g_1)^2$ vanishes at $F_{12}=0$.
Thus the Nielsen-Olesen tachyon in supergravity triggers the 
decay to flat space.

It is clear from the orbifold examples in \cite{Adams:2001sv},
that energy
is a good observable to compare in closed string tachyon
condensation. The orbifold examples considered there
show that the localized closed string tachyons 
condense so that energy is
minimized.  This example also falls into this simple pattern.
The fact that this model admits a tachyon 
potential with a fixed point ensures that
it is possible to study time dependent evolution of the tachyon.
During the condensation process the space changes from a space of
constant curvature to flat space, a time dependent study of this
phenomenon would be interesting.

\subsection{Renormalization group flow in supergravity}

In this section we study the RG flow induced by the tachyon in
supergravity and show that the Nielsen-Olesen tachyon triggers the
system to flow in the direction of decreasing field strengths towards
flat space. 
The RG equations are given by
\bea{rgeq}
\frac{\dot{g_{\mu\nu}}}{\alpha'} &=& 
-R_{\mu\nu} - 2 \nabla_\mu\nabla_\nu \Phi
+\frac{1}{4}H_{\mu\alpha\beta} 
H_\nu^{\;\;\alpha\beta} + \frac{\alpha'}{8}
\rm{Tr} (F_{\mu\alpha} F_{\nu}^\alpha), \\ \nonumber
\frac{ \dot{\Phi}} {\alpha' } &=& 
-\frac{1}{2} \nabla^\mu \nabla_\mu \Phi + \partial_\mu\Phi \partial^\mu
\Phi - \frac{1}{24} H_{\mu\nu\rho} H^{\mu\nu\rho} - \frac{\alpha'}{64}
\rm{Tr} ( F_{\mu\nu} F^{\mu\nu}),  \\ \nonumber
\frac{\dot{B}_{\mu\nu}}{\alpha'} &=& \frac{1}{2}\nabla^\alpha
H_{\alpha\mu\nu} - \partial^\alpha\Phi H_{\alpha\mu\nu}, \\ \nonumber
\frac{\dot{A}_\mu}{\alpha'}  
&=& \frac{1}{2} \hat{D}^\alpha F_{\alpha\mu} -
\partial^\alpha\Phi F_{\alpha\mu} +\frac{1}{4} F_{\nu\rho}
H_\mu^{\;\;\nu\rho} 
\eea
The renormalization group equations are complicated, however we can
obtain a consistent set of RG equations for backgrounds
containing small magnetic fields.
We make the 
following ansatz for the solutions of the 
RG flow which is valid to order
$O(\epsilon^2)$
\bea{flowansatz}
ds^2 &=& -(dt + a_i(x^1, x^2) dx^i) ^2 + 
dx^i dx_i + dx^mdx_m \\ \nonumber
a_i &=& \epsilon_{ij} \tilde{f} x^j +   g_i(x^1, x^2), \quad 
B_{ti} =  \epsilon_{ij}\tilde{f} x^j + 
b_i(x^1, x^2) \\ \nonumber
A_i^3 &=& \epsilon_{ij}  Q f x^j +
\phi_i(x^1, x^2), \quad 
A_i^\pm =  W_i^{\pm}(x^1, x^2)
\eea
Here $\phi_i, g_i$ are of the $O(\epsilon^2)$ and
 the off diagonal fluctuations of the gauge bosons 
which contain the tachyon occur at the  $O(\epsilon)$. 
We assume that the background field $f$ to be small and $f\sim
O(\epsilon^2)$. Using this ansatz we show that we will obtain a
consistent set of RG equations up to $O(\epsilon^3)$.
The off diagonal fluctuations are expanded in the complete set of 
functions which 
satisfy the background gauge condition $\hat{D}^i W_i^{\pm} =0$
and are eigen functions of the operator $O_{ij}^\pm$ given in
\eq{operator}.
The functions $\phi_i$ also satisfy the background gauge condition
$\hat{D}^i\phi_i =0$.
In section 4.1 we have  constructed these eigen
functions. An important property of these functions is that
$F_{ij}^{\pm} \sim O(\epsilon^3)$. This can be verified for the lowest
Landau wave functions given in \eq{wavefn1}, and it is easy to see
that it holds for the higher Landau levels.

Whatever initial conditions we start our RG evolution, they must
satisfy the constraint equations of gravity given in \eq{gra-const}
since these are constraints on the initial velocities.
As discussed in section 5.2 
the constraint equations can be satisfied by imposing
$g_i=b_i$ and  \eq{dil-constraint}. If these equations are satisfied
the $00$ component of the RG equation for the metric 
and the RG equation for the dilaton are automatically satisfied.
The remaining RG equations for the metric reduce to
\be{red-RG-met}
\dot{g}_i = \frac{\alpha'}{2} \partial_j G_{ji},
\ee
where we have used $g_i=b_i$ and \eq{dil-constraint}.
For the field strength $G_{12} = -\tilde{f} + 
\partial_1 g_2 -\partial_2 g_1$
the above equation implies
\nonumber
\be{rg-fieldst}
\dot{G}_{12} = \frac{\alpha'}{2} (\partial_1^2 + \partial_2^2)
G_{12}
\ee
The $01$ and the $02$ RG equations for the $B$-field
also reduce to the above set of equations  while the $12$ equations
is satisfied to $O(\epsilon^3)$. The RG equations  for the gauge
field  to $O(\epsilon^3)$ are given by
\bea{gauge-rg}
\dot{\phi}_i &=& \frac{\alpha'}{2}  \partial_j 
F^3_{ji}  
\\ \nonumber
\dot{W}_i^+ &=& \frac{\alpha'}{2} \left[ 
O_{ij}^+ W_{j}^+ -i W_j^+\partial_j\phi_i + i
\phi_j\partial_j W_i^+ - i W_j^+(\partial_j \phi_i -\partial_i\phi_j)
\right. \\ \nonumber
&+& \left. 2 W_j^+(W_j^+W_i^- - W_i^-W_j^+)\right] \\ \nonumber
\dot{W}_i^- &=& \frac{\alpha'}{2} \left[ 
O_{ij}^- W_{j}^+ +i W_j^-\partial_j\phi_i - i
\phi_j\partial_j W_i^- + i W_j^-(\partial_j \phi_i -\partial_i\phi_j)
\right. \\ \nonumber
&+& \left. 2 W_j^-(W_j^-W_i^+ - W_i^+W_j^-)\right]
\eea
Here 
$F^3_{ij} = -Qf\epsilon_{ij} 
+ \partial_{i} \phi_{j} - \partial_{j} \phi_{i} +
2i(W_i^+ W_j^- - W_j^-W_j^+)$. 
The RG equation for the dilaton relates the metric with the
gauge field, to $O(\epsilon^3)$   
this constraint is given by
\be{dil-rg}
G_{12} = \sqrt{\frac{\alpha'}{2}}\frac{1}{Q}
F^3_{12}
\ee
From the RG equation for the diagonal component of the gauge field we
obtain the following equation
\be{rg-gaugst}
\dot{F}^3_{12} = \frac{\alpha'}{2} (\partial_1^2 + \partial_1^2)
F^3_{12}
\ee
Thus we 
see that the RG equation for the field strength constructed from the 
metric  \eq{rg-fieldst} and the RG
equation for the diagonal component of the gauge field strength are
consistent to $O(\epsilon^3)$. 
The study of the RG flow has now reduced to the flows of
the gauge field
and the above set of equations can be studied numerically by
expanding $W_i^\pm$ in the eigen functions of the operator
$O_{ij}^\pm$.

However, it is possible to 
show that there is a fixed point with vanishing field strength and
curvature. 
The method of obtaining this solution is identical to that done just
for the gauge field in the section  2.2. There we saw that the
equations for the gauge field has a fixed point at 
$\phi_i =0$ with all the coefficients of  the 
Landau levels in $W_i^\pm$ vanishing except the tachyon. The RG
equations for $W_1^+$ reduces to
\be{tac-red-rg}
\dot{W}_1^+ = \frac{\alpha'}{2} \left( Qf W_1^+ - 4 W_1 |W_1^+|^2
\right)
\ee
Thus at the fixed point
the tachyon gets an expectation value $|W_1^+|^2 = Qf/4$, the other
components of the tachyon can be obtained from the relations in 
\eq{rel-llg}.  This point
corresponds vanishing of the field strengths $F_{12}^\pm$ and
$F_{12}^3$. From \eq{dil-rg} we see that $G_{12}$ also vanishes. As the
curvature of this space is proportional to $G_{12}^2$, we see that the
RG flow set up by the tachyon ends up in flat space. It is easy to see
from the RG equations for the field strength that expectation value for
the tachyon drives the RG flow to decreasing curvatures towards flat
space.

To study renormalization group flow induced by
localized  tachyons on  orbifolds, 
an entropy corresponding to localized closed string states
called $g_{cl}$ was defined in \cite{Harvey:2001wm}, 
and it was conjectured that $g_{cl}$ decreases along the RG flow. 
It has been shown in \cite{Basu:2002jt} that the
gauge degrees of freedom do not contribute to $g_{cl}$. In the
system with magnetic field from the $SO(32)$ group of the heterotic
string the localized tachyons belong to the
gauge degrees of freedom. Therefore at least these states are not
taken into account by $g_{cl}$. It would be interesting to compute
$g_{cl}$ for the case with the magnetic field from the $SO(32)$
gauge group and check if it decreases along the RG flow set up by the
tachyon. Another physically motivated quantity which can decrease
along RG flow is the energy. In  all the orbifold examples of
\cite{Adams:2001sv}, the energy of the intial orifold is always higher
than that of the end point. For the system studied in this paper it is
clear that the energy functional given in \eq{energy-func} decreases
along the RG flow as field strengths decrease along RG flow.

\section{Stability against pair production}

Constant fields in a theory are usually unstable to decay via pair
production of particles which are are charged with respect to that
field. For instance, a constant electric field in $U(1)$ gauge theory
decays by pair production of electron-positron pairs
\cite{Schwinger:1951nm}. 
One might expect the backgrounds  studied in this paper also might be
unstable to decay via pair production of particles. In this section we
show that these backgrounds are  stable with respect to decay
by particle creation. 
The reason can be traced to the fact that there is no particle
creation in a electromagnetic plane wave \cite{Schwinger:1951nm}.
All the invariants constructed from the electromagnetic field strength
of the plane wave vanish and thus the effective action describing the
vacuum polarization vanish. A similar argument was used in \cite{Deser}
to argue that there is no particle creation in a gravitational wave.
We use the same method to show that there is no particle creation in
the uniform magnetic field backgrounds considered in this paper.

It is convenient to first study particle creation in the background
given in \eq{back-bos}. The effective action describing vacuum
polarization in this background is obtained by integrating virtual
particles charged with respect to the background fields. Whatever the
character of the virtual particles involved the effective Lagrangian
will be an invariant constructed out of the curvature tensor
$R_{\mu\nu\rho\sigma}$ and the field strength $H_{\mu\nu\rho}$ and
their derivatives.

The effective Lagrangian contains terms of the type
\bea{lag-cur}
L &\sim& R\ldots \nabla R\ldots \nabla^2  R\ldots  + 
H\ldots \nabla H \ldots \nabla^2 H \ldots   \\ \nonumber
&+& R\ldots H \nabla R \ldots 
\nabla^2 R \ldots \nabla H \ldots \nabla^2 H \ldots
\eea
From the properties of the plane wave metric \eq{back-bos} given in
appendix B. it is easy to see  
that the covariant indices of all quantities, the curvature, the
metric, the connection and the field strength $H$ contain only the
indices $u$ and $i$. While the contraviant indices of all
quantities contain only the indices $v$, $i$. Thus any invariant
constructed out of these quantities has to vanish. Note that terms which
involve covariant derivatives of $H$ and  
the curvature also have covariant
indices $u$ and $i$ and contravariant indices $v$ and $i$, 
therefore any invariant constructed out of
covariant derivatives also vanish. Thus the effective Lagrangian
describing vacuum polarization vanishes and there is no pair
production in these backgrounds.
Due to this reason the effective action describing vacuum polarization
for the background \eq{kk-bos} which is obtained
from the plane wave type background \eq{back-bos} by Kaluza-Klein
reduction also vanishes. In summary we see that there is no pair
production of particles in the backgrounds 
studied in this paper. It will be interesting to check this conclusion
by performing an explicit 1-loop calculation in string theory
analogous to \cite{Bachas:1992bh}. 
The
amplitude for pair production can be extracted from the imaginary part
of the 1-loop amplitude.

\section{Conclusions}

We have studied tachyon condensation in backgrounds containing uniform
magnetic fields in heterotic string theory. 
When the magnetic field is embedded in the $SO(32)$ group of the
the heterotic string it is possible to study the tachyonic mode
corresponding to the Nielsen-Olesen instability within supergravity. 
The tachyon can be identified as a fluctuation of the supergravity
fields.  We constructed an energy functional and 
evaluated the closed string 
tachyon potential, the minimum of the tachyon potential corresponds to
flat space.
We studied the world sheet renormalization group flow in the
supergravity approximation. For small values of the magnetic field
we have obtained 
up a consistent set of renormalization group equations. 
We showed that the RG flow set up by the tachyon drives the background
to flat space.
This implies that  tachyon signals the
instability of the uniform magnetic field background to decay to flat
space. 
Thus this system with localized closed string tachyon 
falls into the general pattern that, 
localized closed string tachyons tend to 
decay to flat space.

This system is of further interest as it admits a dual description
in type I theory. In type I theory the tachyons studied
in the heterotic string are in  the open string sector. 
In open string tachyon condensation the boundary entropy or the
tension of the branes involved is a quantity which decreases along 
RG flow set up by the tachyon. It would be interesting to find out
what the boundary entropy would be for the 
type I background considered in this paper. 
It is natural to 
expect that boundary entropy on the type I side will be given
by the tachyon potential evaluated in this paper.

Finally, most systems which are unstable due to perturbative tachyons
are also unstable non-perturbatively due to tunneling 
and the end points are usually same
For example the
D-brane anti D-brane systems are unstable to formation of a bounce
eating up the branes \cite{Banks:1995ch,Callan:1998kz} 
and in the Melvin background the 
perturbative process of tachyon condensation and the non-perturbative
process involving brane nucleation leads to the same end point 
\cite{Gutperle:2002bp}.
On the other hand the systems studied in this paper are stable 
with respect to decay by pair production particles. Essentially this
is because these systems are Kaluza-Klein reduction of a
special classes of plane waves.

\acknowledgments
I would like to thank Alex Buchel, Jaume Gomis, Gary Horowitz, 
Kengo Maeda, Shiraz
Minwalla, Joe Polchinski, Radu Roiban and Ashoke Sen 
for useful discussions and suggestions.
I thank Ashoke Sen for a careful reading of an earlier version of this
manuscript.
I am grateful for hospitality at 
the Caltech-USC center; Centre for Theoretical Studies, Bangalore; 
Harish-Chandra Research Institute, Allahabad; 
Institute for Mathematical Sciences,  Chennai and 
Tata Institute for Fundamental Research, Mumbai during various stages of
this work and for 
fruitful interactions with the high energy group at these
institutes. The work of the author is supported by NSF grant
PHY00-98395.

\appendix
\section{The tachyon potential for the Nielsen-Olesen
instability}
In this appendix we provide the details in evaluating the tachyon
potential. We substitute the expansion given in \eq{fluct} into the
Yang-Mills action and obtain an effective action for the tachyon.
We will organize the terms in the potential according to the order in
fluctuations.
The quadratic term in the fluctuation is given by
\be{quad}
-\frac{1}{4g_{\rm{YM}}^2 } 
 \int d^4x \rm{Tr}\left( -2 \delta A_\nu 
\left( D_\mu D^\mu \delta A^\nu - 2i [F^{\mu\nu}, \delta A_\mu ]
\right) \right)
\ee
Substituting the fluctuation in \eq{fluct} in \eq{quad} we obtain the
following quadratic terms
\be{quad2}
S_2=
-\frac{1}{ 4g_{\rm{YM}}^2 } 
\int d^2 x\left(
- 16 \sqrt{l_1l_2} f Q|\chi|^2
+ 4\int d^2k
\frac{l_1l_2}{4\pi^2} \phi(-k) k^2 \phi(k) \right)
\ee
here the integral is over $x^0$ and $x^3$, we have performed the
integral over $x^1$ and $x^2$.
The cubic term in fluctuations is given by
\be{cubic}
-\frac{1}{ 4g_{\rm{YM}}^2 } 
\int d^4 x \rm{Tr}\left( 2i [\delta A_\mu ,\delta_\nu] \{ D^\mu \delta
A^\nu - D^\nu \delta A^\mu \} \right )
\ee
Again substituting the fluctuation from \eq{fluct} in 
the above equation we obtain
\be{cubic2}
S_3 = 
\frac{16}{ 4g_{\rm{YM}}^2 }  \int d^2x 
\int d^2k \left( \frac{l_1l_2}{4\pi^2}
|k| \phi(k) e^{-\frac{k^2}{2f} } |\chi|^2 \right)
\ee
Finally, the quartic term in the fluctuation is given by
\be{quartic}
-\frac{1}{ 4g_{\rm{YM}}^2 } 
\int d^4x -\rm{Tr}
\left([\delta A_\mu , \delta A_\nu][\delta A^\mu, \delta A^\nu] 
\right)
\ee
Substituting \eq{fluct} in \eq{quartic} we get
\be{quartic2}
S_4 = 16 \frac{l_1 l_2fQ}{\pi} |\chi|^4
\ee
The total action in these fluctuations is given by $S= S_2 + S_3 +
S_4$. We can eliminate the $\phi(k)$ using its classical equation of
motion. This gives the following  
\be{scalars}
\phi(-k) = 2
\frac{e^{-\frac{k^2}{2f}} }{|k|} |\chi|^2
\ee
Now substituting this value of $\phi$ in the action we obtain the
following effective action for the tachyon 
\be{t-pot}
S= -\frac{16}{4g_{YM}^2} \int d^2 x -fQ\sqrt{l_1l_2} |\chi|^2 +
\frac{f Ql_1 l_2}{2\pi}|\chi|^4
\ee

\section{Properties of the plane wave metric}

We list useful properties of the plane wave metric 
given in \eq{back-bos}
The metric and inverse metric components are given by
\bea{met-inmet}
g_{uv} = \frac{1}{2},\quad g_{iu} = \frac{a_i}{2},\quad g_{ij} =
\delta{ij}, \\ \nonumber
g^{uv} = 2, \quad g^{vv} = a_1^2 + a_2^2, \quad g^{vi} = -a_i, \quad
g^{ij} = \delta^{ij}.
\eea
The Cristoffel symbols are given by
\be{cristofel}
\Gamma^i_{uj} = \frac{1}{4} \epsilon_{ij} \tilde{f}, \quad
\Gamma^{v}_{ui} = \frac{1}{4} \epsilon_{ij} a_j \tilde{f}. 
\ee
Finally the components of the curvature tensor and the B-field are
given by
\be{pp-cur}
R_{iuju} = -\delta_{ij} \frac{\tilde{f}^2}{16}, \quad H_{jiu} =
\epsilon_{ij} \tilde{f}
\ee

\bibliographystyle{utphys}
\bibliography{het}
\end{document}